\documentclass[twocolumn]{article}
\usepackage{sample,epsfig,colortbl}

\definecolor{Gray}{gray}{0.90}

\def\BibTeX{{\rm B\kern-.05em{\sc i\kern-.025em b}\kern-.08em
    T\kern-.1667em\lower.7ex\hbox{E}\kern-.125emX}}

\title{Use of Wikipedia Categories in Entity Ranking}

\author{{\em James A. Thom}\\[1ex]
        RMIT University\\Melbourne, Australia\\[1ex]
        {james.thom@rmit.edu.au} \and
        {\em \hspace{0.3in} Jovan Pehcevski \hspace{0.5in} Anne-Marie Vercoustre}\\[1ex]
        INRIA\\Rocquencourt, France\\[1ex]
        {\{jovan.pehcevski, anne-marie.vercoustre\}@inria.fr}}
\date{}

\begin{document}

\maketitle
\thispagestyle{empty}

        \begin{figure}[b]
	~\\
        \noindent
        {\small\raggedright\bf
        Proceedings of the 12th Australasian 
	Document Computing Symposium,
	Melbourne, Australia,
        December 10, 2007.
	Copyright of this article remains with the authors.
        }
        \end{figure}

\paragraph*{Abstract}
%IEEE allows italicized abstract
\noindent
{\it 
Wikipedia is a useful source of knowledge that has many applications in
language processing and knowledge representation.
The Wikipedia category graph can be compared with the class hierarchy in
an ontology; it has some characteristics in common as well as some differences.
In this paper, we present our approach for answering entity ranking queries 
from the Wikipedia. 
In particular, we explore how to make use of Wikipedia categories to improve entity ranking effectiveness. 
Our experiments show that using categories of example entities works significantly better than using loosely defined target categories.
}

%\paragraph*{Keywords} 
%Information Retrieval, Web Documents, Digital Libraries

\section{Introduction}
%Wikipedia is a good source of knowledge that can be used, and has
%been used both as ontology and as a corpus.
%We want to do more, we want to make use of the structure of the Wikipedia to
%assist particular information retrieval tasks.
%Google often highly ranks Wikipedia pages.
%But we are interested in different retrieval acitivity, \emph{entity ranking}
%(which is a little bit similar to
%Google Sets\footnote{\url{http://labs.google.com/sets}}).

%Information systems contain references to many named entities.
%In a well-structured database system it is exactly clear what are references to named entities,
%whereas in semi-structured information sources (such as web pages) it is harder to identify them within a text.
%An entity could be, for example, an organisation, a person, a location, or a date. 
%Because of the importance of named entities, several very active and related research areas have emerged in recent years,
%including: entity extraction/tagging from texts,
%entity reference solving (e.g. ``The president of the Republic''), entity disambiguation (e.g. which Michael Jackson),
%question-answering, % (who/when/where/what),
%expert search, and entity ranking (also known as entity retrieval). % as answers to a query.

Semi-structured text documents contain references to many named entities but, unlike fields
in a well-structured database system, it is hard to identify the named entities within text.
An entity could be, for example, an organisation, a person, a location, or a date. 
Because of the importance of named entities, several very active and related research areas have emerged in recent years,
including: entity extraction/tagging from texts,
entity reference solving (e.g. ``The president of the Republic''), entity disambiguation (e.g. which Michael Jackson),
question-answering, % (who/when/where/what),
expert search, and entity ranking (also known as entity retrieval). % as answers to a query.

Entity ranking is very different from the related problem
of entity extraction.
The objective of \emph{entity extraction} is to identify named entities
from plain text and tag each and every occurrence; whereas the objective of
\emph{entity ranking} is to search for entities in a semi-structured collection 
%The goal of {\em entity ranking} is to retrieve entities as answers to a query.
%The objective is not to tag the names of the entities in documents but rather to
and to get back a list of the relevant entity names as answers to a query 
(with possibly a page or some description associated with each entity). 

The Initiative for the Evaluation of XML retrieval (INEX) is running a new track on entity ranking in 2007~\cite{vries:ent07}, using Wikipedia as its document collection.
In Wikipedia, pages correspond to entities which are organised into (or attached to) categories.
References to entities and their categories
occur frequently in natural language. 
For example, ``France'' is an named entity that corresponds to the Wikipedia page about France, 
belonging to categories such as ``European Countries'' and ``Republics''.

%An example of an entity ranking query is: ``European countries where I can pay
%with Euros''. 
%If we use Wikipedia as our source for entities, the answer to this entity
%ranking query will be a list of entities (Wikipedia pages) corresponding to European countries
%that use the Euro as their currency.

%The Initiative for the Evaluation of XML retrieval (INEX) has proposed a new track on entity ranking~\cite{vries:ent07}, using Wikipedia as its document collection.
%Two tasks are proposed for the INEX 2007 entity ranking track:
%a task where the category of the expected entity answers is provided; and
%a task where a few (two or three) of the expected entity answers are provided.
%The inclusion of target categories (in the first task) and example entities (in the second task)
%makes these quite different tasks from the task of full-text retrieval,
%and the combination of the query and example entities (in the second task) makes it a task quite different from the task addressed by an application such as Google Sets\footnote{http://labs.google.com/sets} where only entity examples are provided. 

There are two tasks in the INEX 2007 entity ranking track:
a task where the category of the expected entity answers is provided; and
a task where a few (two or three) of the expected entity answers are provided.
The inclusion of target categories (in the first task) and example entities (in the second task)
makes these quite different tasks from the task of full-text retrieval,
and the combination of the query and example entities (in the second task) makes it a task quite different from the task addressed by an application such as Google Sets\footnote{http://labs.google.com/sets} where only entity examples are provided. 

In this paper, we present our approach to entity ranking that augments the initial full-text information retrieval approach
with information based on hypertext links and Wikipedia categories. 
In our previous work we have shown the benefits of using categories in entity ranking compared to full-text retrieval~\cite{SAC08}. 
Here we particularly focus on how best to use the Wikipedia category information to improve entity ranking. 

%In the next section we review related work %on using links in ranking and
%on similarity of ontology classes and discuss how (or whether) it could be applied to Wikipedia categories. 
%In Section~\ref{sec:INEX} we describe
%the two tasks that will be carried out by the INEX 2007 entity ranking track. 
%In Section~\ref{sec:wikipedia} we describe the XML version of Wikipedia used by INEX and 
%analyse some of its features. 
%In Section~\ref{sec:category} we present various category similarity approaches we consider for the two INEX 2007 entity ranking tasks, while in Section~\ref{sec:approach} we provide a global overview of our entity ranking approach.
%Our experimental results are presented in Section~\ref{sec:experiments}. 
%The final section lists our conclusions and outlines future work directions.

\section{Related Work}

The traditional entity extraction problem lies in the ability to extract named entities from plain text using natural language processing techniques or statistical methods and intensive training from large collections. 
Benchmarks for evaluation of entity extraction have been performed for the Message Understanding Conference (MUC)~\cite{Sund:muc91} and for the Automatic Content Extraction (ACE) program~\cite{ace:2006}. 
%In that context, training is done on a large number of examples in order to identify extraction patterns (rules).
%and to apply them in the tagging phase.
%The goal is to eventually tag those entities and use the tag names to support future information retrieval.
%However, in the context of large collections such as the web or Wikipedia, it is not possible, nor even desirable,
%to tag in advance all the entities in the collection,
%although many occurrences of named entities in the text may be used as anchor text for sources of hypertext links.
%Instead, since we are dealing with semi-structured documents (HTML or XML), we could exploit the explicit document structure to infer effective extraction patterns or algorithms.
%Recent research in named entity extraction has developed approaches that are not language dependant 
%and do not require lots of linguistic knowledge.

\subsubsection*{Entity extraction}

McNamee and Mayfield~\cite{mcna:coling02} developed a system for entity extraction 
based on training on a large set of very low level textual patterns found in tokens.
Their main objective was to identify entities in multilingual texts and classify them into 
one of four classes (location, person, organisation, or ``others''). 
Cucerzan and Yarowsky~\cite{cuce:emnlp99} describe
and evaluate a language-independent bootstrapping algorithm based on iterative learning and re-estimation of contextual and morphological patterns. 
It achieves competitive performance when trained on a very short labelled name list.

Other approaches for entity extraction are based on the use of external resources, such as an ontology or a dictionary.
Popov et al.~\cite{popo:iswc03} use a populated ontology for entity extraction, 
while Cohen and Sarawagi~\cite{cohe:kdd04} exploit a dictionary for named entity extraction. 
Tenier et al.~\cite{teni:egc06} use an ontology for automatic semantic annotation of web pages. 
Their system first identifies the syntactic structure that characterises an entity in a page. 
It then uses subsumption to identify the more specific concept for this entity, 
%Their approach is based on learning techniques, similar to wrapping techniques, 
combined with reasoning exploiting the formal structure of the ontology.

\subsubsection*{Using ontology for entity disambiguation}

Hassell et al.~\cite{hass:iswc06} use a ``populated ontology'' to assist in disambiguation of entities,
for example names of authors using their published papers or domain of interest. 
They use text proximity between entities to disambiguate names
(e.g. organisation name would be close to author's name).
They also use text co-occurrence, for example for topics relevant to an author.
%They mention that in some cases they need segmentation of the text to limit the context of the co-occurrence.
%They use the notion of ``popular entity'' associated to a given relationship (for example, an author is popular if it has many ``authored'' links). They also use semantic relationships such as co-authoring.
So their algorithm is tuned for their actual ontology,
while our algorithm is more based on the structural properties of the Wikipedia.
%(there is not much semantics on the category relations or page links).

Cucerzan~\cite{cuce:emnlp07} uses Wikipedia data for named entity disambiguation.
He first pre-processed a version of the Wikipedia collection (September 2006),
and extracted more than 1.4 millions entities with an average of 2.4 surface forms by entities.
He also extracted more than one million (entities, category) pairs that were further filtered out to 540 thousand pairs.
Lexico-syntactic patterns, such as titles, links, paragraphs and lists, are used to build co-references of entities
in limited contexts. 
%However, the overwhelming number of contexts that could be extracted this way requires the use of heuristics to limit the context extraction.
The knowledge extracted from Wikipedia is then used for improving entity disambiguation in the context of web and news search. 

\subsubsection*{Ontology similarity}

Since Wikipedia has some but not all characteristics associated
with an ontology, one could think of adapting some of the similarity measures proposed for comparing concepts
in an ontology and use those for comparing categories in Wikipedia.
Ehrig et al.~\cite{ehrig:ecis05} and
Blanchard et al.~\cite{blan:interop05} have surveyed various such similarity
measures.
These measures are mostly reflexive and symmetric~\cite{ehrig:ecis05} and 
take into account the distance (in the path) between the concepts,
the depth from the root of the ontology and the common ancestor of the concepts, and the density of concepts on the paths between the concepts and from
the root of the ontology~\cite{blan:interop05}.

All these measures rely on a strong hierarchy of the ontology concepts and a subsumption hypothesis in the parent-child relationship.
Since those hypothesis are nor verified in Wikipedia (see Section~\ref{sec:wikipedia}), we could not use those similarity functions. 
Instead we experimented with similarities between sets of categories and lexical similarities between category names.

\subsubsection*{Entity ranking}

Fissaha Adafre et al.~\cite{FissahaAdafre:ranlp07} form entity neighbourhoods for every entity, which are based on clustering of similar Wikipedia pages
using a combination of extracts from text content and following both incoming and outgoing page links.
These entity neighbourhoods are then used as the basis for retrieval for the two entity ranking tasks.

Our approach is similar in that it uses XML structural patterns (links) rather than textual ones to identify potential entities. 
It also relies on the co-location of entity names with some of the entity examples (when provided). 
However, we also make use of the category hierarchy to better match the result entities with the expected class of the entities to retrieve.
 
\section{INEX 2007 XML entity ranking track}
\label{sec:INEX}

%{\bf Short Presentation of the track}
The INEX XML entity ranking track is a new track that is being proposed in 2007.
The track will use the Wikipedia XML document collection 
as its test collection.

%that has been used by various INEX tracks in 2006~\cite{INEX06-Overview}.

Two tasks are planned for the INEX Entity ranking track in 2007~\cite{vries:ent07}:
\begin{description}
\item[task 1:] {\em entity ranking}, which aims to retrieve entities of a given category that satisfy a topic described in natural language text; and
\item[task 2:] {\em list completion}, where given a topic text and a number of examples, the aim is to complete this partial list of answers.
\end{description}

\begin{figure}
%\centering
\small
\hrule
\begin{verbatim}

<inex_topic> 
<title>
European countries where I can pay with Euros
</title>
<description>
I want a list of European countries where
I can pay with Euros.
</description>
<narrative>
Each answer should be the article about a
specific European country that uses the
Euro as currency.
</narrative>
<entities>
   <entity ID="10581">France</entity>
   <entity ID="11867">Germany</entity>
   <entity ID="26667">Spain</entity>
</entities>
<categories>
   <category ID="185">european countries
   </category>
</categories>
</inex_topic>
\end{verbatim}
\hrule
\caption{\small Example INEX 2007 XML entity ranking topic}
\label{fig:topic}
%\vskip -6pt
\end{figure}

An example of an INEX entity ranking topic is shown in Figure~\ref{fig:topic}.
In this example, the {\tt title} field contains the plain content only query, 
the {\tt description} provides a natural language description of the information need,
and the {\tt narrative} provides a detailed explanation of what makes an entity answer relevant.
In addition to these fields, the {\tt entities} field provides a few of the expected entity answers for the topic (task 2),
while the {\tt categories} field provides the category of the expected entity answers (task 1).

%Different tasks within the track are based on being given different
%combinations of:
%\begin{itemize}
%\item Query: What European countries use the Euro?
%\item Example entities: France Germany
%\item Example categories: Countries
%\end{itemize}

%[Relevance judging is simpler than other INEX tasks, as only need to determine whether an article 
%(representing an entity) is correct answer or not.
%The track will explore letting participating groups to vote on what are correct answers.
%A harder task (not for 2007) would be to use NLP to infer the
%entities and categories from the topic.]

\section{Wikipedia XML document collection}
\label{sec:wikipedia}

As Wikipedia is fast growing and evolving it is not possible to use the actual online Wikipedia for experiments,
and so there is a need to use a stable collection to do evaluation experiments
that can be compared over time.
Denoyer and Gallinari~\cite{Wikipedia} have developed an XML-based corpus based on a snapshot of the Wikipedia, 
which has been used by various INEX tracks in 2006 and 2007. %~\cite{INEX06-Overview}.
It differs from the real Wikipedia in some respects (size, document format, category tables), but it is a very realistic approximation.

\subsubsection*{Entities in Wikipedia}

In Wikipedia, an {\em entity} is generally associated with an article (a Wikipedia page) describing this entity.
For example, there is a page for every country, most famous people or organisations, places to visit, and so forth.
In Wikipedia nearly everything can be seen as an entity with an associated page. 
%Wikipedia pages have unique names that correspond to
%\emph{entities} (e.g. France, Germany).
The Wikipedia is also a rich source of links,
though some are to pages that do not exist yet!
When mentioning entities in a new Wikipedia article, authors are encouraged to link at least 
the first occurrence of the entity name to the page describing this entity.
This is an important feature as it allows to easily locate potential entities, 
which is a major issue in entity extraction from plain text. 
 
For example, in the small extract from Euro page shown in
Figure~\ref{fig:europage}, there are 18 links 
(shown as underlined) to other pages in the Wikipedia, of which 15 are links to {\em countries}.

\begin{figure}
%\centering
\hrule
\vspace{3mm}
``The {\bf euro} \ldots
%(\underline{currency sign}: \underline{E} ; \underline{banking code}: {\bf EUR})
is the official \underline{currency} of the \underline{Eurozone} (also known as the Euro Area), which consists of the \underline{European} states of \underline{Austria}, \underline{Belgium}, \underline{Finland}, \underline{France}, \underline{Germany}, \underline{Greece}, \underline{Ireland}, \underline{Italy}, \underline{Luxembourg}, \underline{the Netherlands}, \underline{Portugal}, \underline{Slovenia} and \underline{Spain}, and will extend to include \underline{Cyprus} and \underline{Malta} from 1 January 2008.''
\vspace{1mm}
\hrule
\caption{\small Extract from the Euro Wikipedia page}
\label{fig:europage}
%\vskip -6pt
\end{figure} 

\subsubsection*{Categories in Wikipedia}

Wikipedia also offers categories that authors can associate with Wikipedia pages.
There are 113,483 categories in the INEX Wikipedia XML collection, which are organised in a graph of categories.
Each page can be associated with many categories (2.28 as an average).

Wikipedia categories have unique names (e.g. ``France'', ``European Countries''). %, and can have both multiple sub-categories and multiple parent-categories.
New categories can also be created by authors, although they have to follow Wikipedia recommendations 
in both creating new categories and associating them with pages. 
For example, the Spain page is associated with the following categories: 
``Spain'', ``European Union member states'', ``Spanish-speaking countries'', ``Constitutional monarchies'' 
(and some other Wikipedia administrative categories).

%A category can have sub-categories and parent categories. 

Some properties of Wikipedia categories include:
\begin{itemize}
\item	a category may have many subcategories and parent categories;
\item	some categories have many associated pages (i.e. large {\em extension}), while others have smaller extension;
\item	a page that belongs to a given category extension generally does not belong to its ancestors' extension; for example, the page Spain does not belong to the category ``European countries'';
\item	the sub-category relation is not always a subsumption relationship; for example, ``Maps of Europe'' is a sub-category of ``Europe'', but the two categories are not in an {\em is-a} relationship; and
\item	there are cycles in the category graph.
\end{itemize}
Yu et al.~\cite{yu:cikm07} explore these properties in more detail.

\section{Category similarity approaches}
\label{sec:category}

To make use of the Wikipedia categories in entity ranking,
we define similarity functions between:
\begin{itemize}
\item categories of answer entities and target categories (for task 1), and
\item categories of answer entities and a set of categories attached to the entity examples (for task 2).
\end{itemize}

\subsubsection*{Task 1} 

We first define a very basic similarity function that computes the ratio of common categories between 
the set of categories  
$\mathsf{cat}(t)$ 
associated to an answer entity page $t$ and 
the set $\mathsf{cat}(C)$ which is
the union of the provided target categories $C$:

\begin{equation}\displaystyle
	S_C(t) = \frac{| \mathsf{cat}(t) \cap \mathsf{cat}(C)|}{ | \mathsf{cat}(C) | }
\end{equation}

The target categories will be generally very broad, so it is to be expected that the answer entities would not generally belong to these broad categories. Accordingly, we defined several extensions of the set of categories, both for the target categories and the categories attached to answer entities. 

The extensions are based on using sub-categories and parent categories in the graph of Wikipedia categories.
We define $\mathsf{cat_{d}}(C)$ as the set containing the target category and its sub-categories (one level down) 
and $\mathsf{cat_{u}}(t)$ as the set containing the categories attached to an answer entity $t$ and their parent categories (one level up). 
Similarity function can then be defined using the same ratio as above except that $\mathsf{cat}(t)$ is replaced 
with $\mathsf{cat_{u}}(t)$ and 
$\mathsf{cat}(C)$ with $\mathsf{cat_{d}}(C)$.
 
Another approach is to use lexical similarity between categories.
For example, ``european countries'' is lexically similar to ``countries'' since they both contain the word ``countries'' in their names. 
We use an information retrieval approach to retrieve similar categories: 
we have indexed all the categories using their names as corresponding documents.  
By sending the category names C as a query to the search engine, we then retrieve all the categories that are lexically similar to C. 
   
We keep the top M ranked categories and add them to C to form the set $\mathsf{Ccat}(C)$. 
We then use the same similarity function as before, where $\mathsf{cat}(C)$ is replaced with $\mathsf{Ccat}(C)$. 
We also experiment with two alternative approaches: by sending the title of the topic T as a query to the search engine (denoted as $\mathsf{Tcat}(C)$); and by sending both the title of the topic T and the category names C as a query to the search engine (denoted as $\mathsf{TCcat}(C)$).

An alternative approach of using lexical similarity between categories is to index the categories using their names and the names of all their attached entities as corresponding documents. For example,  if C=``countries'', the retrieved set of categories $\mathsf{Ccat}(C)$ may contain not only the categories that contain ``countries'' in their names, but also categories attached to entities with names lexically similar to ``countries''.

\subsubsection*{Task 2} 

In task 2, the categories attached to entity examples are likely to correspond to very specific categories, 
just like those attached to the answer entities. 
We define a similarity function that computes the ratio of common categories between 
the set of categories attached to an answer entity page $\mathsf{cat}(t)$ and 
the set of the union of the categories attached to entity examples $\mathsf{cat}(E)$:

%We can use as a reasonable baseline the very basic similarity function that computes the ratio of common categories between 
%the set of categories associated to the target page $\mathsf{cat}(t)$ and 
%the set of the union of the provided examples $\mathsf{cat}(E)$:

\begin{equation}\displaystyle
	S_C(t) = \frac{| \mathsf{cat}(t) \cap \mathsf{cat}(E)|}{ | \mathsf{cat}(E) | }
\end{equation}

We also expand the two sets of categories by adding the parent categories to calculate $\mathsf{cat_{u}}(t)$ and $\mathsf{cat_{u}}(E)$ and apply the same similarity function as above.

\section{Our approach to entity ranking}
\label{sec:approach}

Our approach to identifying and ranking entities combines: 
(1) the full-text similarity of the entity page with the query; 
(2) the similarity of the page's categories with the target categories or the categories of the entity examples; and 
(3) the links to a page from the top ranked pages returned by a search engine for the query.

\subsection{Architecture}

The approach involves several modules and functions that are used for
processing a query, submitting it to the search engine, applying our entity
ranking algorithms, and finally returning a ranked list of entities. 
We use Zettair\footnote{http://www.seg.rmit.edu.au/zettair/} as our choice for a full-text search engine. 
Zettair is a full-text information retrieval system developed by RMIT, which returns pages ranked by their similarity score to the query. 
We used the Okapi BM25 similarity measure that has proved to work well on the INEX~2006 Wikipedia test collection~\cite{Iskandar-INEX06}.

Our approach involves the following modules:

\begin{itemize}
%\item the topic module takes an INEX topic as input (as the topic example shown in Figure~\ref{fig:topic})
%and generates the corresponding Zettair query and the list of entity examples (as one option, 
%the example entities may be added to the query);

\item The search module sends the query to Zettair and returns a list of scored Wikipedia pages. The assumption is that a good entity page is a page that answers the query. % (or the query extended with examples).

\item The link extraction module extracts the links from a selected number of highly ranked pages,\footnote{We discarded external links and some internal collection links that do not refer to existing pages in the INEX Wikipedia collection.} 
together with the information about the paths of the links (XML paths). The assumption is that a good entity page is a page that is referred to by a page answering the query; this is an adaptation of the Google PageRank~\cite{Google} and HITS~\cite{Klein99} algorithms to the problem of entity ranking. 

\item The linkrank module calculates a weight for a page based (among other things) 
on the number of links to this page (see~\ref{linkrank}). The assumption is that a good entity page is a page that is referred to from contexts with many occurrences of the entity examples. A coarse context would be the full page that contains the entity examples. Smaller and better contexts may be elements such as paragraphs, lists, or tables~\cite{ECIR08}.

\item The category similarity module calculates a weight for a page based on the similarity of the page categories 
with that of the target categories or the categories attached to the entity examples (see~\ref{category}). The assumption is that a good entity page is a page associated with a category close to the target categories or categories of the entity examples.

\item The full-text retrieval module calculates a weight for a page based on its initial Zettair score (see~\ref{zet}).

\end{itemize}

The global score for a page is calculated as a linear combination of three normalised scores coming out of the last 
three modules (see~\ref{global}).

%The overall process for entity ranking is shown in Figure~\ref{process}.
The above architecture provides a general framework for evaluating entity ranking.
%
%since it supports replacing some modules by more advanced modules (e.g. the
%category similarity module), or by providing a more efficient implementation of a module.
%It also includes an evaluation module to assist in tuning the system by
%varying the parameters and to globally evaluate the entity ranking approach.

\subsection{LinkRank score}
\label{linkrank}

The linkrank function calculates a score for a page, based on the number of
links to this page, from the first N pages returned by the search engine in
response to the query. We carried out some experiments with different
values of N and found that N=20 was an acceptable compromise between performance
and discovering more potentially good entities.

We use a very basic linkrank function that, for an answer entity page $t$ that is pointed to by a page $p$, takes into account the Zettair score of the referring page $z(p)$, and the number of reference links to the answer entity page $\#links(p,t)$:

\begin{equation}\displaystyle
	S_L(t) = \sum_{r=1}^N z(p_r) * f(\#links(p_r,t))
\end{equation}

\noindent where $f(x) = x$ (when there is no reference link to the answer entity page, $f(x) = 0$). 

The linkrank function can be implemented in a variety of ways; 
for task 2 where entity examples are provided, we have also experimented by weighting
pages containing a number of entity examples, or by exploiting the locality of links around the entity examples~\cite{ECIR08}. 
This more complex implementation of the linkrank function is outside the scope of this paper. %, since here we are  focused on improving the category similarity function by comparing the effectiveness of our entity ranking approach across the two tasks.

\subsection{Category score}
\label{category}

The basic category score $S_C(t)$ is calculated for the two tasks as follows:

{\bf task 1} 
\begin{equation}\displaystyle
	S_C(t) = \frac{| \mathsf{cat}(t) \cap \mathsf{cat}(C)|}{ | \mathsf{cat}(C) | }
\end{equation}

{\bf task 2} 

\begin{equation}\displaystyle
	S_C(t) = \frac{| \mathsf{cat}(t) \cap \mathsf{cat}(E)|}{ | \mathsf{cat}(E) | }
\end{equation}

We then consider variations on the category score $S_C(t)$ given the considerations in Section~\ref{sec:category},
using various combinations of replacing $\mathsf{cat}(t)$ with $\mathsf{cat_{u}}(t)$,
replacing $\mathsf{cat}(C)$ with $\mathsf{cat_{d}}(C)$, $\mathsf{Ccat}(C)$, $\mathsf{Tcat}(C)$ or $\mathsf{TCcat}(C)$, 
and replacing $\mathsf{cat}(E)$ with $\mathsf{cat_{u}}(E)$.

\begin{table*}[tp]
\caption{\small Performance scores for runs using different retrieval strategies in our category similarity module ($\alpha$0.0--$\beta$1.0), obtained for task 1 by different evaluation measures. For the three runs using lexical similarity, the Zettair index comprises documents containing category names (C), or documents containing category names and names of entities associated with the category (CE). The number of category answers retrieved by Zettair is M=10. For each measure, the best performing score is shown in bold.}
\begin{minipage}{8cm}
\begin{center}
\begin{tabular}{l c c c c}
\hline \hline
 & \multicolumn{2}{c}{P[r]} & & \\
 \cline{2-3}
 Run & 5 & 10 & R-prec & MAP  \\
\hline \hline
$\mathsf{cat}(C)$-$\mathsf{cat}(t)$ & 0.229 & 0.250 & 0.215 & 0.196 \\
$\mathsf{cat_d}(C)$-$\mathsf{cat_u}(t)$ & 0.243 & 0.246 & 0.209 & 0.185 \\
\hline
$\mathsf{Ccat}(C)$-$\mathsf{cat}(t)$ & 0.214 & 0.250 & 0.214 & 0.197 \\
$\mathsf{Tcat}(C)$-$\mathsf{cat}(t)$ & {\bf 0.264} & 0.261 & 0.239 & 0.216 \\
$\mathsf{TCcat}(C)$-$\mathsf{cat}(t)$ & {\bf 0.264} & {\bf 0.286} & {\bf 0.247} & {\bf 0.226} \\
\hline \hline
\end{tabular}
(C) Index of category names
\end{center}
\end{minipage}
~
\begin{minipage}{8cm}
\begin{center}
\begin{tabular}{l c c c c}
\hline \hline
 & \multicolumn{2}{c}{P[r]} & & \\
 \cline{2-3}
 Run & 5 & 10 & R-prec & MAP  \\
\hline \hline
$\mathsf{cat}(C)$-$\mathsf{cat}(t)$ & 0.229 & {\bf 0.250} & {\bf 0.215} & {\bf 0.196} \\
$\mathsf{cat_d}(C)$-$\mathsf{cat_u}(t)$ & {\bf 0.243} & 0.246 & 0.209 & 0.185 \\
\hline
$\mathsf{Ccat}(C)$-$\mathsf{cat}(t)$ & 0.157 & 0.171 & 0.149 & 0.148 \\
$\mathsf{Tcat}(C)$-$\mathsf{cat}(t)$ & 0.171 & 0.182 & 0.170 & 0.157 \\
$\mathsf{TCcat}(C)$-$\mathsf{cat}(t)$ & 0.207 & 0.214 & 0.175 & 0.173 \\
\hline \hline
\end{tabular}
(CE) Index of category and entity names
\end{center}
\end{minipage}
\label{C-CE-runs}
\end{table*}

\subsection{Z score}
\label{zet}

The Z score assigns the initial Zettair score to an answer entity page. If the answer page does not appear in the final list of ranked pages returned by Zettair, then its Z score is zero:
\begin{equation}
	S_Z(t) = \left\{ \begin{array}{ll}
	z(t) & \mbox{if page }t \mbox{ was returned by Zettair} \\
&\\
0 & \mbox{otherwise }
\end{array}
\right.
\end{equation}

\subsection{Global score}
\label{global}

The global score $S(t)$ for an answer entity page is calculated as a linear combination of three normalised scores, the linkrank score $S_L(t)$, the category similarity score $S_C(t)$, and the Z score $S_Z(t)$:
\begin{equation}
	S(t) = \alpha S_L(t)  + \beta S_C(t) + (1 -  \alpha - \beta) S_Z(t)
\end{equation}
where $\alpha$ and $\beta$ are parameters that can be tuned.

A special case of interest here is when only the category score is used ($\alpha=0.0$, $\beta=1.0$).
It allows us to evaluate the effectiveness of various category similarity functions and the overall benefit of using categories.

\section{Experimental results}
\label{sec:experiments}

We now present results that investigate the effectiveness of our entity ranking approach for the two entity ranking tasks. We start by describing the test collection we developed for entity ranking.

\subsection{Test collection}

There is no existing set of topics with relevance assessments for entity ranking, although such a set will be developed for the INEX XML entity ranking track in~2007. 
So for these experiments we developed our own test collection based on a selection of topics from the INEX~2006 ad hoc track.
We chose 27 topics that we considered were of an ``entity ranking'' nature, 
where for each page that had been assessed as containing relevant information, we reassessed whether or not it was an entity answer, 
and whether it {\em loosely} belonged to a category of entities we had {\em loosely} identified as being the target of the topic. 
If there were entity examples mentioned in the original topic these were usually used as entity examples in the entity topic.
Otherwise, a selected number (typically 2 or 3) of entity examples were chosen somewhat arbitrarily from the relevance assessments.
To this set of 27 topics we also added the Euro topic example (shown in Figure~\ref{fig:topic}) 
that we had created by hand from the original INEX description of the entity ranking track~\cite{vries:ent07}, resulting in total of 28 entity ranking topics.

We use mean average precision (MAP) as our primary method of evaluation, but also report results using several alternative information retrieval measures: mean of P[5] and P[10] (mean precision at top 5 or 10 entities returned), and mean R-precision (R-precision for a topic is the P[R], where R is the number of entities that have been judged relevant for the topic). When dealing with entity ranking, the ultimate goal is to retrieve all the answer entities at the top of the ranking. Although we believe that MAP may be more suitable than the other measures in capturing these aspects, part of the track at INEX 2007 will involve determining what is the most suitable measure.

\subsection{Task 1}

%Compare the following
%\begin{itemize}
%\item Query index (of category names) with \verb+<title>+ 
%\item Query index (of category names with entity names) with \verb+<title>+ 
%\item Query index (of category names) with \verb+<title>+ and \verb+<categories>+
%\item Query index (of category names with entity names) with \verb+<title>+ and \verb+<categories>+
%\item Query index (of category names) with \verb+<categories>+
%\item Query index (of category names with entity names) with \verb+<categories>+
%\item Just use \verb+<categories>+ (BASELINE)
%\item Expand \verb+<categories>+ with up/down of both query category(s) and the
%categories of the returned entities (HOW MANY LEVELS - may need to go at least 5?)
%\end{itemize}

For this task we carried out three separate investigations. First, we wanted to investigate the effectiveness of our category similarity module when varying the extensions of the set of categories attached to both the target categories and the answer entities. We also investigated the impact that this variation had on the effectiveness when the two different category indexes are used by Zettair. Second, for the best category similarity approach we investigated the optimal value for the parameter M (the number of category answers retrieved by Zettair). Last, for the best category similarity approach using the optimal M value, we investigated the optimal values for the $\alpha$ and $\beta$ parameters. The aim of this investigation is to find the best score that could be achieved by our entity ranking approach for task 1.

\subsubsection*{Investigating category similarity approaches}

The results of these investigations are shown in Tables~\ref{C-CE-runs}(C) and~\ref{C-CE-runs}(CE).\footnote{The first two runs do not use any of Zettair's category indexes and are included for comparison.} 
Several observations can be drawn from these results.

First, the choice of using the Zettair category index can dramatically influence the entity ranking performance. When cross-comparing the results in the two tables, we observe that the three lexical similarity runs using the Zettair index of category names substantially outperform the corresponding runs using the Zettair index of category and entity names. The differences in performance are all statistically significant ($p < 0.05$). Second, the run that uses the query that combines the terms from the title and the category fields of an INEX topic ($\mathsf{TCcat}(C)$-$\mathsf{cat}(t)$) performs the best among the three runs using lexical similarity, and overall it also performs the best among the five runs when using the Zettair index of category names. However, the differences in performance between this and the other four runs are not statistically significant. Third, extending the set of categories attached to both the target categories and the answer entities overall does not result in an improved performance, although there are some (non-significant) early precision improvements.

\subsubsection*{Investigating the parameter M}

The above results show that the best effectiveness for our category similarity module ($\alpha$0.0--$\beta$1.0) is achieved when using the Zettair index of category names, together with the query strategy that combines the terms from the title and the category fields of an INEX topic. For these experiments we used a fixed value M=10 for the parameter M that represents the number of category answers retrieved by Zettair. However, since this was an arbitrary choice we also investigated whether a different value of M could also have a positive impact on the retrieval effectiveness. We therefore varied M from 1 to~20 in steps of 1, and measured the MAP scores achieved by our best performing $\mathsf{TCcat}(C)$-$\mathsf{cat}(t)$ run using the Zettair index of category names.

\begin{figure}
\begin{center}
\epsfxsize=7.5cm
\centering\epsfbox{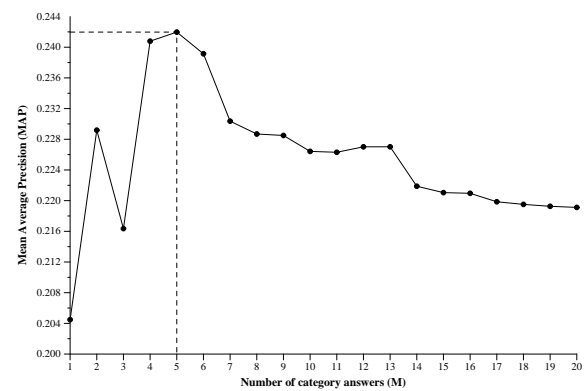}
\end{center}
\caption
{\small Investigating the optimal value for the number of category answers retrieved by Zettair, when using the run $\mathsf{TCcat}(C)$-$\mathsf{cat}(t)$.}
\label{fig-M-tuning}
\end{figure}

Figure~\ref{fig-M-tuning} shows the results of this investigation. We observe that a value of 5 for the parameter M yields the highest MAP score (0.242) for our category similarity module, which is a 7\% relative performance improvement over the MAP score obtained with M=10. This performance improvement is statistically significant ($p < 0.05$).

\subsubsection*{Investigating the combining parameters $\alpha$ and $\beta$}

To find the best score that could be achieved by our entity ranking approach for task 1, we used the run $\mathsf{TCcat}(C)$-$\mathsf{cat}(t)$ with the optimal value M=5 and investigated various combinations of scores obtained from the three modules. 
We calculated MAP over the 28 topics in our test collection, as we varied $\alpha$ from 0 to 1 in steps of 0.1. For each value of $\alpha$, we also varied $\beta$ from 0 to $(1 - \alpha)$ in steps of 0.1. %Table~\ref{Q-MAP-training} shows these results. 
We found that the highest MAP score (0.287) is achieved for $\alpha=0.1$ and $\beta=0.8$. This is a 19\% relative performance improvement over the best score achieved by using only the category module ($\alpha$0.0--$\beta$1.0). This performance improvement is statistically significant ($p < 0.05$). We also calculated the scores using mean R-precision instead of MAP as our evaluation measure, and we again observed the same performance behaviour and optimal values for the two parameters. 

\subsection{Task 2}

%Compare the following
%\begin{itemize}
%\item Current system\footnote{possibly with better linkrank from ECIR paper - alternatively we leave combining the best of both to the SIGIR and/or journal paper} on 27 topics (BASLINE) with $\beta = 1$, $beta = 0.3$, $beta = 0.6$
%\item go up (to parents of) categories of example entites
%\item go up (to parents of) categories of answer entities
%\item go up (to parents of) both categories of example entites and categories of answer entities 
%\end{itemize}

For this task we carried out two separate investigations. First, as with task 1 we wanted to investigate the effectiveness of our category similarity module when varying the extensions of the set of categories attached to both the example and the answer entities. Second, for the best category similarity approach we investigated the optimal values for the $\alpha$ and $\beta$ parameters, with the aim of finding the best score that could be achieved by our entity ranking approach for task 2. 
%Third, we compared the category similarity approaches across the two tasks to investigate which of the two query strategies (target categories or example entities) is better for entity ranking.

\begin{table}[tp]
\caption{\small Performance scores for runs using different retrieval strategies in our category similarity module ($\alpha$0.0--$\beta$1.0), obtained for task 2 by different evaluation measures. For each measure, the best performing score is shown in bold.}
\begin{center}
\begin{tabular}{l c c c c}
\hline \hline
 & \multicolumn{2}{c}{P[r]} & & \\
 \cline{2-3}
 Run & 5 & 10 & R-prec & MAP  \\
\hline \hline
$\mathsf{cat}(E)$-$\mathsf{cat}(t)$ & {\bf 0.536} & {\bf 0.393} & {\bf 0.332} & {\bf 0.338} \\
\hline
$\mathsf{cat}(E)$-$\mathsf{cat_u}(t)$ & 0.493 & 0.361 & 0.294 & 0.313 \\
$\mathsf{cat_u}(E)$-$\mathsf{cat}(t)$ & 0.407 & 0.336 & 0.275 & 0.255 \\
$\mathsf{cat_u}(E)$-$\mathsf{cat_u}(t)$ & 0.357 & 0.332 & 0.269 & 0.261 \\
\hline \hline
\end{tabular}
\end{center}
\label{Task2-runs}
\end{table}

\subsubsection*{Investigating category similarity approaches}

The results of these investigations are shown in Table~\ref{Task2-runs}. We observe that, as with task 1, extending the set of categories attached to either (or both) of the example and answer entities does not result in an improved performance. %More specifically, we observe that, irrespective of the evaluation measure used, the run $\mathsf{cat}(E)$-$\mathsf{cat}(t)$, which uses only the set of categories directly attached to both the example and the answer entities, substantially outperforms the other three runs that use various extensions of the two sets of categories. 
The differences in performance between the best performing run that does not use the extended category sets and the other three runs that use any (or both) of these sets are all statistically significant ($p < 0.05$).

\subsubsection*{Investigating the combining parameters $\alpha$ and $\beta$}

To find the best score that could be achieved by our entity ranking approach for task 2, we used the run $\mathsf{cat}(E)$-$\mathsf{cat}(t)$ and investigated various combinations of scores obtained from the three modules. 
We calculated MAP over the 28 topics in our test collection, as we used the 66 combined values for parameters $\alpha$ and~$\beta$. We found that the highest MAP score (0.396) was again achieved for $\alpha=0.1$ and $\beta=0.8$. This score is a 17\% relative performance improvement over the best score achieved by using only the category module ($\alpha$0.0--$\beta$1.0). The performance improvement is statistically significant ($p < 0.05$). 
%We also calculated the scores using mean R-precision instead of MAP as our evaluation measure, and we again observed the same performance behaviour and optimal values for the two parameters. 

\subsection{Comparing Task 1 and Task 2}

To investigate which of the two query strategies (target categories or example entities) is more effective for entity ranking, we compared the scores of the best performing runs across the two tasks. Table~\ref{Task-comparing-runs} shows the results of this comparison, when separately taking into account two distinct cases: a case when using scores coming out of the category module only ($\alpha$0.0--$\beta$1.0); and a case when using optimal global scores coming out of the three modules ($\alpha$0.1--$\beta$0.8). 

We observe that, irrespective of whether category or global scores are used by our entity ranking approach, the run that uses the set of categories attached to example entities (task 2) substantially outperforms the run that uses the set of categories identified by Zettair using the topic title and the target categories (task 1). The differences in performance between the two runs are statistically significant ($p < 0.05$). This finding shows that using example entities is much more effective query strategy than using the loosely defined target categories, which allows for the answer entities to be identified and ranked more accurately.

\section{Conclusions and future work}

\begin{table}[tp]
\caption{\small Comparing best performing runs for task 1 and task 2 for two distinct cases (using either category or global scores). The number of category answers retrieved by Zettair for run $\mathsf{TCcat}(C)$-$\mathsf{cat}(t)$ is M=5. For each case, the best results are shown in bold.}
\begin{center}
\begin{tabular}{l c c c c}
\hline \hline
 & \multicolumn{2}{c}{P[r]} & & \\
 \cline{2-3}
 Run & 5 & 10 & R-prec & MAP  \\
\hline \hline
\multicolumn{5}{l}{ {\bf Category score: $\alpha$0.0--$\beta$1.0} } \\
$\mathsf{TCcat}(C)$-$\mathsf{cat}(t)$ & 0.307 & 0.318 & 0.263 & 0.242 \\
$\mathsf{cat}(E)$-$\mathsf{cat}(t)$ & {\bf 0.536} & {\bf 0.393} & {\bf 0.332} & {\bf 0.338} \\
\hline
\multicolumn{5}{l}{ {\bf Global score: $\alpha$0.1--$\beta$0.8} } \\
$\mathsf{TCcat}(C)$-$\mathsf{cat}(t)$ & 0.379 & 0.361 & 0.338 & 0.287 \\
$\mathsf{cat}(E)$-$\mathsf{cat}(t)$ & {\bf 0.607} & {\bf 0.457} & {\bf 0.412} & {\bf 0.396} \\
\hline \hline
\end{tabular}
\end{center}
\label{Task-comparing-runs}
\end{table}

In this paper, we have presented our entity ranking approach for the INEX Wikipedia XML document collection. 
% which is based on exploiting the interesting semantic properties of the collection. 
We focused on different entity ranking strategies that can be used by our category similarity module, and evaluated these strategies on the two entity ranking tasks. For task 1, we demonstrated that using lexical similarity between category names results in an effective entity ranking approach, so long as the category index comprises documents containing only category names. For task 2, we demonstrated that the best approach is the one that uses the sets of categories directly attached to both the example and the answer entities, and that using various extensions of these two sets significantly decreases the entity ranking performance. For the two tasks, combining scores coming out of the three modules %in our entity ranking approach 
significantly improves the performance compared to that achieved when using scores only from the category module. Importantly, when comparing the scores of the best performing runs across the two tasks, we found that the query strategy that uses example entities to identify the set of target categories is significantly more effective than the strategy that uses the set of loosely defined target categories.

In the future, we plan to further improve the global score of our entity ranking approach by using a better linkrank function that exploits different (static and dynamic) contexts identified around the entity examples. Preliminary results demonstrate that the locality of links around entity examples can indeed be exploited to significantly improve the entity ranking performance compared to the performance achieved when using the full page context~\cite{ECIR08}. To improve the category similarity function, we plan to introduce different category weighting rules that we hope would better distinguish the answer entities that are more similar to the entity examples. We will also be participating
in the INEX~2007 entity ranking track, which we expect would enable us to test our approach using a larger set of topics.

\paragraph*{Acknowledgements}

Part of this work was completed while James Thom was visiting INRIA in 2007.

\begin{small}
\bibliography{sample}
\end{small}

\end{document}